\documentclass[aps,prb,twocolumn,groupedaddress,floatfix,eqsecnum,showpacs,showkeys]{revtex4}
\usepackage{amssymb,amsfonts,amsmath}
\usepackage[dvips]{graphics,color,floatflt,epsfig} 
\usepackage{graphicx} 
\usepackage{psfrag} 
\begin{document}
\newcommand{\Sref}[1]{Section~\ref{#1}}
\newcommand{\sref}[1]{Sec.~\ref{#1}}
\newcommand{\fref}[1]{Fig.~\ref{#1}}
\newcommand{\frefs}[1]{Figs.~\ref{#1}}
\newcommand{\Fref}[1]{Figure~\ref{#1}}
\newcommand{\Frefs}[1]{Figures~\ref{#1}}
\newcommand{\tql}{\textquotedblleft} 
\newcommand{\tqr}{\textquotedblright~}
\newcommand{\tqrc}{\textquotedblright}  

\title{Designer disordered materials with large complete photonic band gaps }

\author{Marian Florescu$^{1}$} 

\email[Electronic Address:]{florescu@princeton.edu}

\author{Salvatore Torquato$^{2,3}$}

\author{Paul J. Steinhardt$^{2,3}$}

\affiliation{$^1$ Department of Physics, Princeton University, Princeton, New Jersey,
  08544, USA}

\affiliation{$^2$Department of Chemistry, Princeton University, Princeton, New Jersey
  08544, USA}

\affiliation{$^3$Princeton Center for Theoretical Sciences, Princeton University,
  Princeton, New Jersey 08544, USA}

\begin{abstract} 
We present designs of \textcolor{black}{2D} isotropic, disordered photonic materials of
arbitrary size with {\it complete} band gaps blocking all directions and polarizations.
The designs with the largest gaps are obtained by a constrained optimization method that
starts from a hyperuniform disordered point pattern, an array of points whose number
variance within a spherical sampling window grows more slowly than the volume.  We argue
that hyperuniformity, combined with uniform local topology and short-range geometric
order, can explain how complete photonic band gaps are possible without long-range
translational order.  We note the ramifications for electronic and phononic band gaps in
disordered materials.
\end{abstract}

\maketitle

\section{Introduction} 
Since their introduction in 1987, photonic band gap materials \cite{c1,c2} have evolved
dramatically and their unusual properties have led to diverse applications, including
efficient radiation sources \cite{e3}, sensors \cite{e4}, and optical computer chips
\cite{e5}. To date, though, the only known large scale dielectric heterostructures with
sizeable, complete band gaps ($\Delta\omega/\omega_C\ge10\%$, say, where $\Delta\omega$ is
the width of the band gap and $\omega_C$ is the midpoint frequency) have been periodic,
which limits the rotational symmetry and defect properties critical for controlling the
flow of light in applications.
\begin{figure}[ht]
{\centerline{ \includegraphics*[angle=0,width=\linewidth]{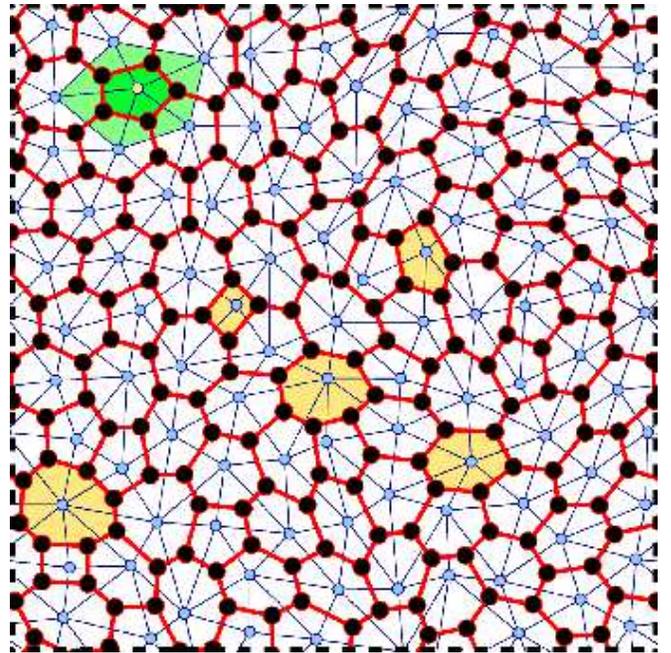}}}
\caption{\label{fig1} Protocol for mapping point patterns into tessellations for photonic
  structure design (see \sref{design}).  First, a chosen point pattern (open circles) is
  partitioned using a Delaunay triangulation (thin lines).  Next, the centroids of the
  neighboring triangles (solid circles) of a given point are connected, generating cells
  (thick lines) around each point, as shown for the five (green) Delaunay triangles in the
  upper left corner of the figure.}
\end{figure}
In this paper, we show that it is possible to design 2D isotropic, translationally
disordered photonic materials of arbitrary size with large complete PBGs. The designs have
been generated through a protocol that can be used to construct different types of
disordered {\it hyperuniform} structures \textcolor{black}{in two or more dimensions},
which are distinguished by their suppressed density fluctuations on long length scales
\cite{c5} and may serve as templates for designer materials with various other novel
physical properties, including electronic, phononic, elastic and transport behavior.

\begin{figure*}[ht]
{\centerline{ \includegraphics*[angle=0,width=1\linewidth]{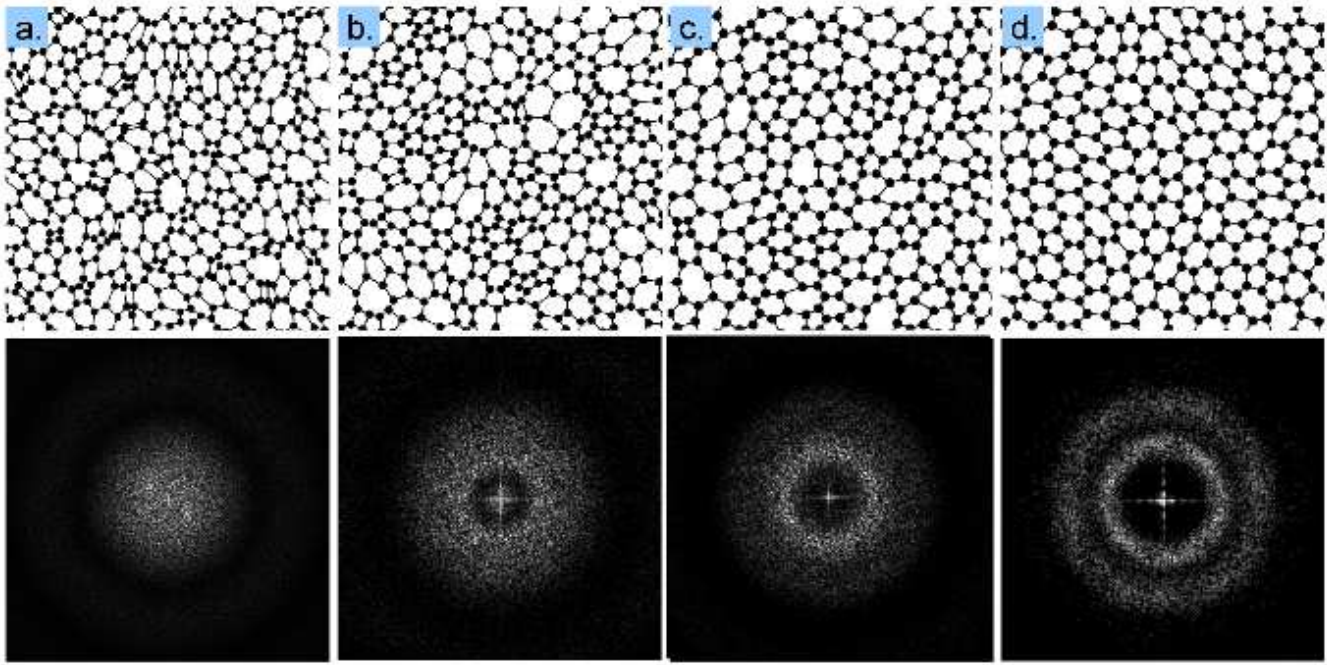}}}
\caption{\label{fig2} Four designs of isotropic dielectric heterostructures derived using
  the protocol in \fref{fig1} and their structure factors, $S(\mathbf{k})$.  Only the
  \fref{fig2}d exhibits a complete PBG.  \fref{fig2}a is a disordered network design
  derived from a Poisson ($p=2$, non-hyperuniform) point pattern.  \fref{fig2}b shows a
  network derived from a nearly hyperuniform equi-luminous point pattern in which the
  structure factor $S(k\rightarrow 0)=S_0>0$ for $k< k_C$.  \fref{fig2}c shows a network
  derived from a RSA point pattern in which the structure factor $S(k\rightarrow 0)>0$ but
  there is more local geometric order than in \fref{fig2}b.  \fref{fig2}d is derived from
  an isotropic, disordered, stealthy hyperuniform pattern with $p=1$ $S(\mathbf{k})$ is
  precisely zero within the inner disk.  Note the two concentric shells of sharply
  increased density just beyond the disk.  These features sharpen as the ordering
  parameter $\chi$ increases; this trend coincides in real space with the exclusion zone
  around each particle increasing and the emergence of complete PBGs.}
\end{figure*}

Here we focus on adapting the protocol for fabricating materials with optimal photonic
properties because of their useful applications and because it is feasible to manufacture
the dielectric heterostructure designs presented in this paper using existing techniques.
Although the goal here is to produce designs for isotropic disordered heterostructures, we
show elsewhere \cite{c8} how the same procedure can be used to obtain photonic
quasicrystals with complete PBGs.  

The design procedure includes a limited number of free parameters (two, in the cases
considered here) that are varied to find the optimal band gap properties.  The
optimization requires modest computational cost as compared with full-blown optimizations
that search over all possible dielectric designs. In practice, although, we find that the
protocol produces band gap properties that are not measurably different from the optima
obtained by brute force methods in cases where those computations have been performed.  To
compute the band gaps for the various disordered structures, we employ a supercell
approximation in which the disordered structure is treated as if it repeats periodically.
We then perform systematic convergence tests to insure that results converge as the
supercell size increases.

Obtaining complete PBGs in dielectric materials without long-range order is
counterintuitive.  We suggest on the basis of a combination of theoretical arguments and
numerical simulations that the PBGs may be explained in the limit large dielectric
constant ratio by a combination of hyperuniformity, uniform local topology, and short-
range geometric order.  All of these conditions are automatically satisfied by photonic
crystals and by all the disordered heterostructures (and quasicrystals) with complete PBGs
produced by our protocol.

We particularly want to emphasize the role of hyperuniformity.  The concept of
hyperuniformity was first introduced as an order metric for ranking point patterns
according to their local density fluctuations \cite{c5}. A point pattern is hyperuniform
if the number variance $\sigma^2(R)\equiv \langle N^2_R\rangle -\langle N_R\rangle^2$
within a spherical sampling window of radius $R$ (in $d$ dimensions) grows more slowly
than the window volume for large $R$, i.e., more slowly than $R^{d}$.  The hyperuniform
patterns considered in this paper are 2D and restricted to the subclass in which the
number variance grows like the window surface area for large $R$, i.e., $\sigma^2(R)=AR$.
The coefficient $A$ measures the degree of hyperuniformity within this subclass: smaller
values of $A$ are more hyperuniform.  In reciprocal space, hyperuniformity corresponds to
having a structure factor $S({\bf k})$ that tends to zero as the wavenumber $k=|{\bf k}|$
tends to zero (omitting forward scattering), i.e., infinite wavelength density
fluctuations vanish.  Hyperuniform patterns include all crystals and quasicrystals, and a
special subset of disordered structures.

Although all crystal and quasicrystal point patterns are hyperuniform, it is considerably
more difficult to identify and/or construct disordered hyperuniform point
patterns. Recently, a collective coordinate approach has been devised to explicitly
produce point patterns with precisely tuned wave scattering characteristics (that is to
say, tuned $S(k)$ for a fixed range of wavenumbers $k$), including a large class of
hyperuniform point patterns, even isotropic, disordered ones \cite{c9}.  Here we apply
these patterns to photonics and present an explicit protocol for designing arrangements of
dielectric materials optimal for photonics from hyperuniform point patterns.  We observe
that there is a strong correlation between the degree of hyperuniformity (smallness of
$A$) for a variety two-dimensional crystal structures as measured in Ref.~\cite{c5} and
the resulting band gaps.  For example, a triangular lattice of parallel cylinders has the
smallest value of $A$ and the largest band gap for light polarized with its electric field
oscillating normal to the plane, whereas a square lattice of cylinders has a larger value
of $A$ and a smaller photonic band gap.  These results motivated us to consider beginning
from seed patterns with a high degree of hyperuniformity to obtain complete PBGs. Indeed,
in the ensuing discussion, we show how this expectation has been explicitly realized in
systematically producing the first known examples of disordered heterostructures of
arbitrary size with complete PBGs.

\begin{figure}[ht]
{\centerline{ \includegraphics*[angle=0,width=\linewidth]{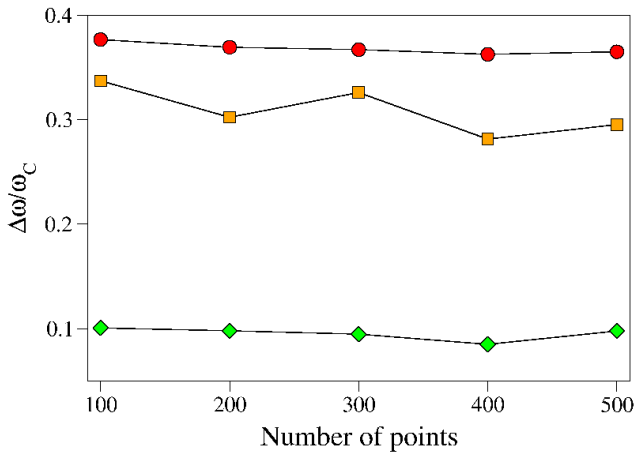}}}
\caption{\label{fig3} Optimal fractional photonic band gaps in photonic structures based
  on stealthy disordered hyperuniform structures of different number of points $N$ with
  $\chi=0.5$. The plot shows that TM- (red circles), TE- (orange squares) and complete-
  (green diamonds) band gaps do not vary significantly with system size.}
\end{figure} 

\section{Design protocol for PBG materials.} 
\label{design}
In the past, photonic crystals displaying large PBGs have been found by a combination of
physical intuition and trial-and-error methods.  Identifying the dielectric decoration
that produces the globally maximal PBG is well known to be a daunting computational task,
despite the recent development of optimization methods, such as gradient-based approaches,
exhaustive search methods, and evolutionary methods \cite{c11,c12,c13}. The major
difficulty in solving this inverse problem comes from the relatively large number of
iterations required to achieve an optimal design and the high computational cost of
obtaining the band structure for complex distributions of dielectric materials, as needed
to simulate heterostructures without long-range order. For instance, the evolutionary
algorithms employed in \cite{preble} require over a 1000 generations of designs to achieve
fully convergence.  (By comparison, our protocol achieves a nearly optimal solution in
only 5-10 iterations.)  Moreover, little progress has been made on rigorous optimization
methods applicable to 3D photonic crystals.

For these reasons, the development of a simpler design protocol that requires vastly less
computational resources is significant.  Because our protocol only optimizes over two
degrees of freedom, it does not guarantee an absolute optimum. However, we find that the
resulting band gap properties are not measurably different from those obtained by the
rigorous optimization methods in the cases where rigorous methods have been
applied. Moreover, our method has already produced the largest known full photonic band
gaps for 2D periodic, quasiperiodic and disordered structures that are too complex for
current rigorous methods to be applied.  The protocol begins with the selection of a point
pattern generated by any means with the rotational symmetry and translational order
desired for the final photonic material.  For crystal, quasicrystal, or random Poisson
patterns, a conventional procedure may be used.  For hyperuniform and other designer point
patterns, we use the previously developed collective coordinate approach \cite{c9} to
produce patterns for certain specific forms of $S(k)$, as described below.

\begin{figure}[ht]
{\centerline{ \includegraphics*[angle=0,width=\linewidth]{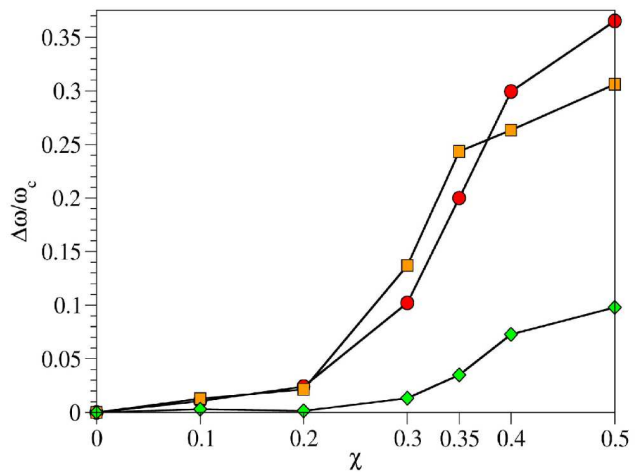}}}
\caption{\label{fig4} A plot showing how the photonic band gap increases as $\chi$, or,
  equivalently, the degree of hyperuniformity and short-range geometric order increases.
  TM (red circles), TE (orange squares) and complete (green diamonds) photonic band gaps
  versus order parameter $\chi$ for disordered stealthy hyperuniform derived using the
  protocol in \fref{fig1}.  The optimal structures have dimensionless cylinder radius
  $r/a=0.189$ for the TM case, dimensionless tessellation wall thickness $w/a=0.101$ for
  the TE case, and ($r/a=0.189$, $w/a=0.031$) for the TM+TE case. }
\end{figure}
\begin{figure}[ht]
{\centerline{\includegraphics*[width=\linewidth]{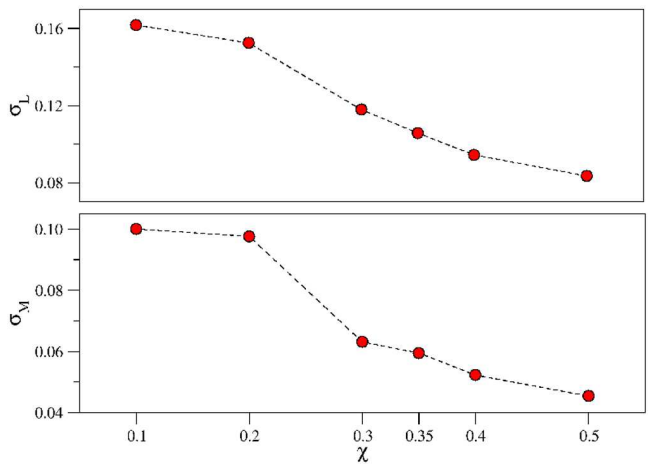}}}
\caption{ \label{fig5} A plot showing how the short-range geometric order for stealth
  disordered structures increases as $\chi$, or, equivalently, the degree of
  hyperuniformity, increases.  (Upper) The standard deviation of the average link length
  vs. $\chi$ for isotropic, disordered, stealthy hyperuniform structures of the type shown
  in \fref{fig2}d. (Lower) The standard deviation of the average link separation
  distribution (calculated as the distance between the mid-points of two neighouring
  links) as a function of $\chi$.  Both plots show a significant decrease in variance as
  $\chi$ increases above 0.35.}
\end{figure}

If the goal were to have a band gap only for TM polarization (electric field oscillating
along the azimuthal direction), the rest of the protocol would simply be to replace each
point in the original point pattern with a circular cylinder and vary the radius of the
cylinders until the structure exhibits a maximum TM band gap \cite{c3,c15,c16,c17}.
However, this design is poor for obtaining a band gap for TE polarization (electric field
oriented in the plane).  We find that the analogous optimum for TE modes is a planar,
continuous trivalent network \cite{c12} (as in the case of the triangular lattice), which
can be obtained from the point pattern using the steps described in \fref{fig1}.  Namely,
construct a Delaunay tiling \cite{c14} from the original two-dimensional point pattern and
follow the steps in \fref{fig1} to transform it into a tessellation of cells.  Then
decorate the cell edges with walls (along the azimuthal direction) of dielectric material
of uniform width $w$ and vary the width of the walls until the maximal TE band gap is
obtained.

Finally, to obtain designs for complete PBGs, the protocol is to optimally overlap the TM
and TE band gaps by decorating the vertices of the trivalent network of cell walls with
circular cylinders (black circles in \fref{fig1}) of radius $r$.  Then, for any given set
of dielectric materials, the maximal complete PBG is achieved by varying the only two free
parameters, $w$ and $r$.  (In practice, the optimal designs obtained by our protocol thus
far have almost the same values of $w$ and $r$ for a given point density, so that a nearly
optimal design may often be achieved without any optimization.)

Although a constrained optimization method like this is not guaranteed to produce the
absolute optimum over all possible designs, in examples where the absolute optimum is
known by rigorous optimization methods \cite{c11,c12,c13}, our protocol produces a design
whose band gap is the same within the numerical error using exponentially less
computational resources.

For the optimization of the two degrees of freedom ($w$ and $r$), the photonic mode
properties must be computed as parameters are varied. Because the computational
requirements are modest, we employ a supercell approximation and use the conventional
plane-wave expansion method \cite{c3,c25} to calculate the photonic band structure; we
generate the disordered pattern within a box of side length $L$ (with periodic boundary
conditions) where $L$ is much greater than the average interparticle spacing and take the
limit as $L$ becomes large.  The PBGs for disordered heterostructures obtained by our
protocol turn out to be equivalent to the fundamental band gap in periodic systems in the
sense that the spectral location of the TM gap, for example, is determined by the resonant
frequencies of the scattering centers \cite{c17} and always occurs between band $N$ and
$N+1$, with $N$ precisely the number of points per unit cell. This behavior underscores
the relevance of the individual scattering center properties on the band gap opening and
can be interpreted in terms of an effective folding of the band structure as a result of
scattering on a collection of $N$ similar (but not necessarily identical) scattering units
distributed hyperuniformly in space.

\section{Results: Disordered photonic materials with large complete PBGs}

To obtain the best results, we consider a subclass of hyperuniform patterns known as {\it
  stealthy}, so-named because they are transparent to incident radiation ($S(k)=0$) for
certain wavenumbers \cite{c9}. In particular, we consider stealthy point patterns with a
structure factor $S(k)$ that is isotropic, continuous and precisely equal to zero for a
finite range of wavenumbers $k <k_C$ for some positive $k_C$. Figure \ref{fig4} presents
four designs of photonic structures derived using the protocol in \fref{fig1} starting
from stealthy point patterns and their structure factors, $S(\mathbf{k})$.

Stealthy hyperuniform patterns are parameterized by $k_C$ or, equivalently, $\chi$, the
fraction of wavenumbers $\mathbf{k}$ within the Brillouin zone that are set to zero; as
$\chi$ increases, $k_C$ and the degree of hyperuniformity increase, thus, decreasing $A$
in our definition of the number variance. When $\chi$ reaches a critical value
$\chi_C$($\approx 0.77$ for 2D systems) the pattern develops long-range translational
order \cite{c9}.

The largest PBGs in hyperuniform patterns occur in the limit of large dielectric contrast;
our band structure computations assume the photonic materials are composed of silicon
(with dielectric constant $\epsilon=11.56$) and air.  To confirm that the computation
converges and the complete PBGs are insensitive to system size, we vary the number of
points per unit cell (sidelength $L$) ranges from $N =100-500$; see \fref{fig3}. For the
purposes of comparison, we use a length scale $a=L/\sqrt{N}$, such that all patterns have
the same point density $1/a^2$.

A significant band gap begins to open for the stealthy hyperuniform designs for
sufficiently large $\chi \approx 0.35$ (but well below $\chi_C$), at a value where there
emerges a finite exclusion zone between neighboring points in the real space hyperuniform
pattern; see \fref{fig4}.  In reciprocal space, this value of $\chi$ corresponds to the
emergence of a range of \tql forbidden\tqr scattering, $S({\mathbf k})=0$ for
$|\mathbf{k}|<k_C$ for some positive $k_C$, surrounded by a circular shell just beyond
$|{\mathbf k}|=k_C$ with increased scattering.  
\begin{figure}[t!]
{\centerline{\includegraphics*[width=0.975\linewidth]{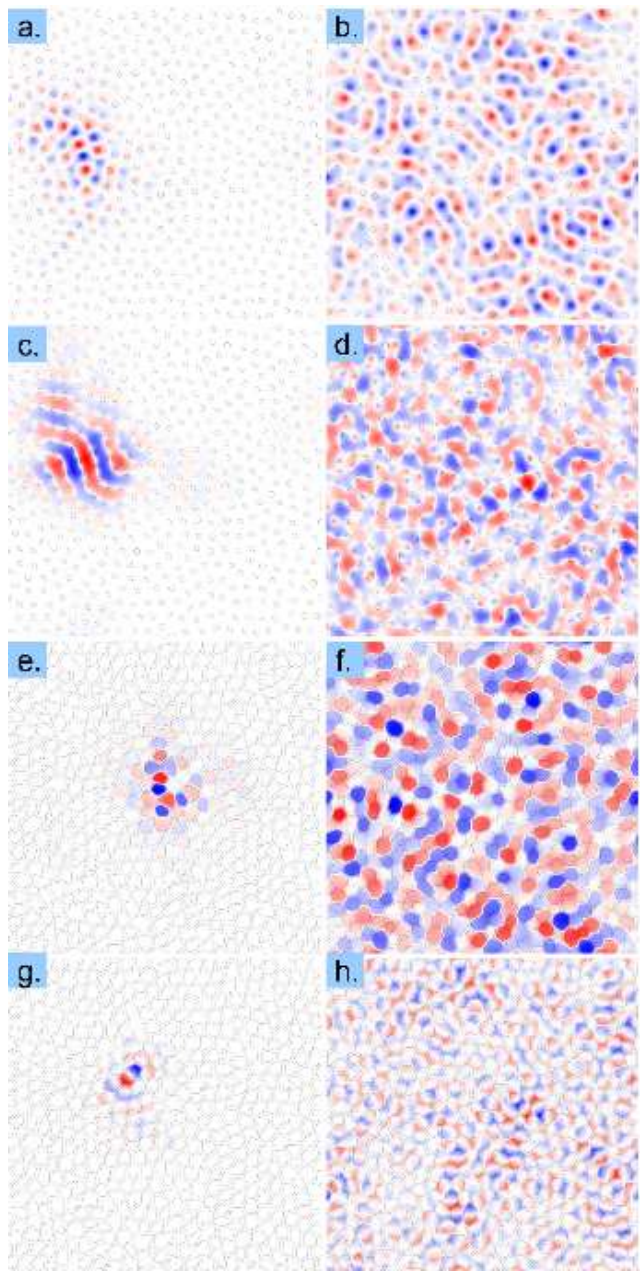}}}
\caption{\label{fig6} (a), (b), (c) and (d): Electric field distribution in hyperuniform
  disordered structures for TM polarization. The structure consists of dielectric
  cylinders ($r/a=0.189$ and $\epsilon=11.56$) in air arranged according to a hyperuniform
  distribution with $\chi=0.5$ and displays a TM PBG of $36.5$\%. (a) Localized and (b)
  extended modes around the lower PBG edge, and (c) localized and (d) extended modes
  around the upper PBG edge.  (e), (f), (g), and (h): Magnetic field distribution in
  hyperuniform disordered structures for TE polarized radiation. The structure consists of
  trihedral network architecture ($w/a=0.101$ and $\epsilon=11.56$) obtained from a
  hyperuniform distribution with $\chi=0.5$ and displays a TE PBG of 31.5\%. (e) Localized
  and (f) extended modes around the lower PBG edge, and (g) localized and (h) extended
  modes around the upper PBG edge.}
\end{figure}
The structures built around stealthy hyperuniform patterns with $\chi =0.5$ are found to
exhibit remarkably large TM (of 36.5\%) and TE (of 29.6\%) PBGs, making them competitive
with many of their periodic and quasiperiodic counterparts.  More importantly, there are
complete PBGs of appreciable magnitude reaching values of about 10\% of the central
frequency for $\chi =0.5$.

Note that the density fluctuations for stealthy patterns are dramatically suppressed for
wavelengths greater than $2\pi/k_C$.  The lower limit $2\pi/k_C$ is directly related to
the midgap frequency $\omega_C$ (see \fref{fig4}) for large enough $\chi$, and the band
width is inversely proportional to the magnitude of the density fluctuations on length
scales greater than $2\pi/k_C$.

A striking feature of the PBGs is their isotropy.  In Ref.~\cite{c13}, an isotropy metric
was introduced that measures the variation in band gap width as a function of incident
angle.  The most isotropic crystal band gap has a variation of 20\%, compared to less than
0.1\% for the hyperuniform disordered pattern in \fref{fig2}d. As noted in our closing
discussion, isotropy can be useful for several applications.

\section{Conditions for photonic band gaps}

Photonic (and electronic) band gaps are commonly associated with long-range translational
order and Bragg scattering, so the examples of disordered PBG materials presented in this
paper are counterintuitive.  We argue below, based on numerical evidence and physical
arguments, that complete photonic band gaps can occur in disordered systems that exhibit a
combination of hyperuniformity, uniform local topology, and short-range geometric order.
This argument has ramifications for electronic and phononic band gaps in disordered
materials, as well.

First, consider the evidence provided by numerical experiments to date.  Photonic crystals
are hyperuniform (an automatic consequence of periodicity) and the known examples with the
largest TM, TE and complete PBGs satisfy the two conditions \cite{c12,c26,souk}.  Our own
numerical experiments indicate that hyperuniformity is a crucial condition.  For example,
we have compared results for the hyperuniform pattern in \fref{fig2}b with networks
generated from non-hyperuniform Poisson point patterns with $p=2$, as in \fref{fig2}a;
{\it equi-luminous} point patterns with $S(k\rightarrow 0) > 0$ for $k < k_C$ where the
non-zero constant $S(0)$ is made very small, as shown in \fref{fig2}c; and with a random-
sequential absorption (RSA) point pattern \cite{c23} generated by randomly, irreversibly
and sequentially placing equal-sized circular disks in a large square box with periodic
boundary conditions subject to a non-overlap constraint until no more can be added.  It
has been shown that such two-dimensional RSA packings have $S(k \rightarrow 0)$ slightly
positive at $k=0$ and increasing as a positive power of $k$ for small $k$ \cite{c24}.  The
latter two patterns are very nearly hyperuniform presenting similar deviations from
hyperuniformity ($S_{e-lum}(k\rightarrow 0)=0.05$ and $S_{RSA}(k\rightarrow 0)=0.053$; and
the RSA network in \fref{fig2}c exhibits uniform topological order (trivalency) and well
defined short-range geometric order; furthermore, these two patterns produce TM and TE
band gaps separately.  Yet none of the three families of patterns has been found to yield
a complete PBG.
  
We also note that hyperuniform stealthy patterns with $\chi< 0.35$ (and keeping all other
parameters fixed) do not produce sizable complete PBGs while those with $\chi>0.35$ do, as
demonstrated in \fref{fig4}.  \fref{fig5} indicates that a difference is the degree of
short-range geometric order, the variance in the near neighbor distribution of link
lengths and the distance between centers of neighboring links.

Based on these and other numerical experiments, we conclude that both hyperuniformity and
short-range geometric order are required to obtain substantial PBGs.  In principle, the
two can be varied independently, but it is notable that patterns with the highest degree
of hyperuniformity also possess the highest degree of short-range geometrical order and
that, for the case of stealthy patterns, both hyperuniformity and short-range geometric
order increase as $\chi$ increases.

To explain how hyperuniformity and short-range geometric order, when combined with uniform
local topology, can lead to a complete PBG, we first return to the point that the band
gaps we have found arise in the limit of large dielectric constant ratio.  In this limit
and for the optimal link widths and cylinder radii, the interaction with electromagnetic
waves is in the Mie scattering limit.  At frequencies near the Mie resonances (which
coincide with the PBG lower band edge frequencies), the scattering of TM electromagnetic
waves in a heterostructure composed of parallel cylinders is similar to the scattering of
electrons by atomic orbitals in cases where the tight-binding approximation can be
reliably applied \cite{c28}.  The same applies for TE modes for any one direction
${\mathbf k}$ if, instead of parallel azimuthal cylinders, there are parallel thick
lines(or walls in the azimuthal direction) in the plane and oriented perpendicular to
${\mathbf k}$; however, to obtain a complete band gap, some compromise must be found to
enable band gap for all directions ${\mathbf k}$.  We conjecture, based on comparison with
rigorous optimization results, our own numerical experiments, and arguments below, that
uniform local topology is advantageous for forming optimal band gaps.  In two dimensions,
this is easiest to achieve in disordered structures without disrupting the short-range
geometric order if the networks are trivalent.

If the arrangement of dielectrics has local geometric order (the variance in link lengths
and inter-link distances is small), the propagation of light in the limit of high
dielectric constant ratio is described by a tight binding model with nearly uniform
coefficients.  In the analogous electronic problem, Weaire and Thorpe \cite{e6} proved
that band gaps can exist in continuous random tetrahedrally coordinated networks, commonly
used as models for amorphous silicon and germanium.  In addition to tight binding with
nearly uniform coefficients, the derivation required uniform tetrahedral
coordination. (Weaire and Thorpe call networks satisfying these conditions {\it
  topologically disordered}.) The analogy in two dimensions is a trivalent
network. Although their proof discussed three dimensions and tetrahedral-coordination
specifically, we find that it can generalize to other dimensions and networks with
different uniform coordination.  Note that our protocol automatically imposes uniform
topology ({\it e.g.}, trivalency in two dimensions) and limits variation of the tight
binding parameters by imposing local geometric order.

To complete their proof, Weaire and Thorpe added a mild stipulation that the density has
bounded variation, defined as the condition that the density remains between two finite
values as the volume is taken to infinity.  This condition is satisfied by any homogeneous
system, hyperuniform or not, and thus is much weaker than hyperuniformity, for which
$\sigma^2(R) = AR$ in the stealth two-dimensional examples.
  
Although bounded variation may be sufficient to obtain a non-zero electronic band gap, we
conjecture that hyperuniform tetrahedrally-coordinated continuous random networks have
substantially larger electronic band gaps than those that do not.  This can be
straightforwardly tested: the collective coordinate method described in Ref.~\cite{c9}
combined with our protocol is a rigorous method for producing hyperuniform (as well as a
range of controlled non-hyperuniform) tetrahedrally-coordinated continuous random network
models.  Our conjecture can be explored by constructing explicit networks and computing
the electronic band gaps.

Analogous questions arise about real amorphous materials made in the laboratory: do
different methods of producing amorphous silicon and germanium result in the same degree
of hyperuniformity and is the behavior of $S(k \rightarrow 0)$ correlated with their
electronic properties?  The same questions apply to phononic properties of disordered
materials.

This line of reasoning also explains why hyperuniformity is more important for obtaining
complete photonic band gaps than electronic band gaps.  For the electronic case, the only
issue is whether there is a gap at all; the width and gap center frequency are not
considered. For the photonic case, a gap is needed simultaneously for both TM and TE, and
the gap centers must have values that allow an overlap.  Also, the goal is not simply to
have a gap, but to have the widest gap possible.  The evidence shows that hyperuniformity
is highly advantageous (perhaps even essential) for meeting these added conditions.

The comparison to electronic band gaps is also useful in comparing states near the band
edges and continuum.  For a perfectly ordered crystal (or photonic crystal), the
electronic (photonic) states at the band edge are propagating such that the electrons
(electromagnetic fields) sample many sites.  If modest disorder is introduced, localized
states begin to fill in the gap so that the states just below and just above are
localized.  Although formally the disordered heterostructures do not have equivalent
propagating states, an analogous phenomenon occurs.  In the upper four panels of
\fref{fig6}, we compare the azimuthal electric field distribution for modes well below or
well above the band gap (upper two panels), which we might call {\it extended} since the
field is distributed among many sites; and then modes at the band edges, which are {\it
  localized}.

We find that the formation of the TM band gap is closely related to the formation of
electromagnetic resonances localized within the dielectric cylinders (\frefs{fig6}(a) and
(b)) and that there is a strong correlation between the scattering properties of the
individual scatterers (dielectric cylinders) and the band gap location.  In particular,
the largest TM gap occurs when the frequency of the first Mie resonance coincides with the
lower edge of the photonic band gap \cite{c17}.  Analogous to the case of periodic
systems, we also find that electric field for the lower band-edge states is well localized
in the cylinders (the high dielectric component), thereby lowering their frequencies; and
the electric field for the upper band-edge states are localized in the air fraction,
increasing their frequencies (see \frefs{fig6}(c) and (d)).  As shown in
\frefs{fig6}(e)-(h), an analogous behavior occurs for the azimuthal magnetic field
distribution for TE modes: for the lower edge state, the azimuthal magnetic field is
mostly localized inside the air fraction and presents nodal planes that pass through the
high-index of refraction fraction of the structure, while the upper edge state displays
the opposite behavior.

The discussion above accounts in a non-rigorous way for the conditions for obtaining PBGs
and all the properties observed in numerical experiments by us and others to date. We hope
to develop the argument into a more precise theory in future work.

\section{Discussion}

This work demonstrates explicitly and proposes an explanation of how it is possible to
design isotropic disordered photonic materials of arbitrary size with complete PBGs.
Although photonic crystals have larger complete band gaps, disordered hyperuniform
heterostructures with substantial complete PBGs offer advantages for many applications.
Disordered heterostructures are isotropic, which is advantageous for use as
highly-efficient isotropic thermal radiation sources \cite{c27} and waveguides with
arbitrary bending angle \cite{c16}. The properties of defects and channels useful for
controlling the flow of light are different for disordered structures. Crystals have a
unique, reproducible band structure; by contrast, the band gaps for the disordered
structures have some modest random variation for different point distributions.  Also,
light with frequencies above or below the band edges are propagating modes that are
transmitted through photonic crystals but are localized modes in the case of 2D
hyperuniform disordered patterns, which give the former advantages in some applications,
such as light sources or radiation harvesting materials.  On the other hand, due to their
compatibility with general boundary constraints, photonic band gap structures based on
disordered hyperuniform patterns can provide a flexible optical insulator platform for
planar optical circuits.  Moreover, eventual flaws that could seriously degrade the
optical characteristics of photonic crystals and perhaps quasicrystals are likely to have
less effect on disordered hyperuniform structures, therefore relaxing fabrication
constraints. The results presented here are obtained for 2D structures, but a direct
extension of our tessellation algorithm to 3D can be used to produces hyperuniform
tetravalent connected network structures. Such a tetravalent connected network decorated
with dielectric cylinders along the its edges could constitute the blueprint for 3D
disordered hyperuniform photonic band gap structures.  (We note that the largest known 3D
photonic band gap is provided by a similar periodic tetravalent network generated from a
diamond lattice \cite{soukT,maldo}). Our preliminary investigation of 3D quasicrystalline
patterns show that the protocol introduced here is able to generate complete photonic band
gaps in 3D quasicrystalline photonic structures, and our plan calls for investigation of
3D hyperuniform disordered structures as well.  Further analysis of the character of the
electromagnetic electromagnetic modes supported by the disordered structures and the
extension to 3D systems may be able to provide a better understanding of the interplay
between disorder and hyperuniformity and between localized and extended electromagnetic
modes in the formation of the photonic band gaps.

Finally, we note that the lessons learned here have broad physical implications.  One is
led to appreciate that all isotropic disordered solids are not the same: as methods of
synthesizing solids and heterostructures advance, it will become possible to produce
different types and degrees of hyperuniformity, and, consequently, many distinct classes
of materials with novel electronic, phononic and photonic properties.

\begin{acknowledgments}
The authors thank R. Batten for generating the hyperuniform and equiluminous disordered
point patterns.  This work was supported by National Science Foundation under Grant No.
DMR-0606415.
\end{acknowledgments}


\end{document}